\documentclass[aps,prb,10pt,
reprint,
showpacs]{revtex4-1}
\usepackage{graphicx}
\usepackage{amsmath}

\renewcommand\Re{\operatorname{Re}}
\renewcommand\Im{\operatorname{Im}}

\begin{document}

\title{Excitation of plasmons in two-dimensional electron gas with defects by microwaves: Wake-field method}

\author{Eduard Takhtamirov$^1$}\email {Electronic address: ed.takhtamirov@gmail.com}
\author{Roderick V. N. Melnik$^{1,2}$}\email {Electronic address: rmelnik@wlu.ca}
\affiliation{$^1$M$^2$NeT Laboratory, Wilfrid Laurier University, Waterloo, Ontario N2L 3C5, Canada\\
$^2$BCAM, Bizkaia Technology Park, 48160 Derio, Spain}


\begin{abstract}
We develop an analytical method to find plasmons generated by microwaves in a two-dimensional electron gas with defects. The excitations are expressed in terms of the wake field of a charged particle moving in plasma. The result explicitly addresses the efficiency of the photon-to-plasmon conversion and the type of excitation. While strong absorption of the radiation by the excitations is reached at larger plasmon wave numbers, intense persistent plasma waves are created at optimal ones. The latter wave numbers depend on the spectrum of plasmons and the distance that the waves are required to travel without being substantially attenuated. Their type, which can be traveling or standing, is governed by the geometry of the defects and the polarization of the radiation. We identify such types of traveling plasmons as circular plasmons, excited at dot defects, and traverse plasmons, excited at straight wire defects and traveling away from (toward) the wires if their group velocity is positive (negative). Nonlinear excitations are also accounted for. In particular, we analyze the zeroth harmonic, linear in the microwave intensity, which has the character of a frozen charge density wave. The interference of elementary wakes from defects arranged in truncated periodic sets can produce amplified plasmons, which are easily portrayed.
\end{abstract}


\maketitle

\section{Introduction}\label{I_intro}

The problem of excitation of plasmons has always received considerable attention associated primarily with experimental verification of theoretical predictions for spectra of quasi-particles in interacting many-electron systems. \cite{Grimes,AllenJr,Theis,Heitmann,Kukushkin1,Kukushkin2} Recently, the interest has been greatly fueled by the development of a new generation of photonic devices, with plasmons playing a key role. \cite{Dyakonov,Allen,Barnes,Shalaev} The efficiency of the photon-to-plasmon conversion is always a major concern. To generate intense plasma waves, one has to deal with a conflict between the requirement to have a system with high electron mobility, beneficial for plasmons, and the necessity to build in a strong enough potential that breaks the system's continuous translational symmetry, inevitably impairing the mobility. Different balanced schemes have to be adopted depending on the specifics of the task. \cite{Kukushkin1,Kukushkin2}

The type of excited plasmon is also an important subject connected to the destination of the excitations, or explicitly to the mechanism of their terminal detection \cite{Muravev} or transformation. \cite{Barnes} If a particular situation requires the excitation of, for example, traveling plasmons, a standing charge density wave should be treated as detrimental rather than just idle because merely the presence of its exciting agent deteriorates the whole system's characteristics. This leads to the efficiency issue again.

For defects arranged in a grating with a spatial period $\delta x$, it was found that light can effectively excite plasmons with wave numbers $Q = 2\pi n/\delta x$, where $n=1,2,\ldots$. \cite{AllenJr,Theis,Heitmann} This gives only a qualitative picture lacking important details. A deeper insight usually requires a direct solution of Maxwell's equations. This can be performed (semi)analytically while considering surface plasmon-polaritons \cite{Fano,Ritchie,Stern_surf,Pitarke} in metals, by using local dielectric constants, \cite{Sondergaard,Rotenberg} or in two-dimensional (2D) semiconductor systems still within the approximation of local conductivities, \cite{Satou} which does not allow description, in particular, of collective excitations with large wave numbers. The Green's-function-based approaches are probably the only means to deal with such situations as the excitation of plasmons by spatially inhomogeneous electromagnetic beams \cite{Sondergaard,Rotenberg} or scattering of the plasmons themselves at defects. \cite{Sondergaard,Bozhevolnyi} But such integral equation methods would invoke very cumbersome calculations with an obscure result if they were applied to systems \cite{Kukushkin2} exhibiting a significant space-time dispersion of the response function.

We develop an effective alternative approach applicable in situations where the homogeneous microwave (MW) radiation acts on an electron gas with defects, which can be treated as weak scatterers. It is analogous to the formalism of the polarization wake of a fast charged particle in plasma. \cite{Neufeld,Echenique} Here it is the static defect, which plays the role of a charged particle, that excites a wake in the electron gas, the latter moving as a whole under the action of the MW field. For simplicity, we put two restrictions on the system under study. First, we regard collective excitations for which the retardation of the electromagnetic interaction can be neglected. In particular, we focus our attention on a 2D electron gas (2DEG) with plasmons \cite{Stern} or magnetoplasmons \cite{Horing,Chiu} (in quantizing magnetic fields) as the lowest-lying excitations. Actually, surface plasmons with large wave numbers, for which the admixture of polaritons is negligibly small, are also inferred. We refer to all these excitations as plasmons unless otherwise stated. And second, we suppose that the internal structure of the defects is not changed by the MW field, which affects only the 2DEG.

The expression we derive below for excited plasmons is then analyzed in terms of the efficiency of the photon-to-plasmon conversion and the type of the resulting excitations. Our results indicate that generally two goals of the conversion should be distinguished from one another: the generation of intense persistent plasma waves and the attainment of strong absorption of the radiation by the excitations. While the former is crucial for fast plasmonic devices, \cite{Dyakonov} the latter is commonly used to study the dispersion of plasmons as such. \cite{Grimes,AllenJr,Theis,Heitmann,Kukushkin1,Kukushkin2} We show that maximum efficiencies of these two processes are generally reached at different plasmon wave numbers and thus different values of the radiation frequency, depending on the spectrum of plasmons.

The paper is organized as follows. In Sec.~\ref{II_EG_with_defects} we review general properties of a 2DEG, the single-electron spectrum of which is parabolic, with embedded defects under the action of MW radiation with a homogeneous electric field. In Sec.~\ref{III_Plasmon_field} we focus on the derived expression for the screened potential of the defects and specify the plasmon contribution. Wakes from a wire defect and a truncated periodic set of wire defects are dealt with in Secs.~\ref{IV_Wake_wire} and \ref{V_Wake_wire_set}, respectively. In Sec.~\ref{VI_Wake_dots} we study wakes from dot defects. In Sec.~\ref{VII_Role_missing} we discuss the effects of retardation and the MW-induced polarization of the defects, neglected in previous sections. Conclusions are given in Sec.~\ref{VIII_conclusion}.

\section{Electron gas with defects in MW field}\label{II_EG_with_defects}

We consider an isotropic 2DEG unbounded in the $(x,y)$ plane of Cartesian coordinates. The system is irradiated by MWs with frequency $\Omega >0$ and in-plane electric field ${\mathbf E}(t) = \left(E_x(t), E_y(t) \right)$ with the amplitudes $\left({\mathcal E}_x, {\mathcal E}_y \right)$:
\begin{equation}\label{field}
E_j(t) = {\mathcal E}_j\, {\mathrm e}^{-i\Omega t} + {\mathcal E}^*_j {\mathrm e}^{i\Omega t}, \quad j=x,y,
\end{equation}
where ${\mathcal E}^*_j$ is the complex conjugate of ${\mathcal E}_j$. The Hamiltonian of a 2D electron with charge $-e< 0$ and effective mass $m^\star$ is
\begin{equation}
H_0 = \frac {\mathbf p^2} {2m^\star}  + e{\mathbf E}(t) {\mathbf r},
\label{ham0}
\end{equation}
where $\mathbf p = (p_x,p_y)$ is the momentum operator, which includes the vector potential of the external magnetic field (if present), and $\mathbf r = (x,y)$ is a 2D coordinate.

It is convenient to change the reference frame to the one associated with the instantaneous position of the classic electron or, in magnetic fields, of the classic center of the electron's cyclotron orbit. For the Schr\"odinger equation, such a transformation, realized with a unitary operator $S$, was proposed more than five decades ago. \cite{Husimi} Recently, it has been used by a number of researchers (see the work by Dmitriev {\em et al.} \cite{Dmitriev} and references therein) in connection with the experimentally observed MW-induced resistance oscillations (see the pioneering work by Zudov {\em et al.}). \cite{Zudov}

Below, we use two properties of this transformation. First, the transformed Hamiltonian $\widetilde H_0$ does not contain the radiation field in this moving reference frame and coincides with the Hamiltonian for the rest reference frame without radiation:
\begin{equation}
\widetilde H_0 = S^\dag H_0 S = \frac {\mathbf p^2} {2m^\star},
\label{tilde_ham0}
\end{equation}
where $S^\dag$ is the Hermitian conjugate of the operator $S$. And second, the transformation is homogeneous, the coordinate being transformed as follows:
\begin{equation}
\widetilde {\mathbf r} = S^\dag {\mathbf r} S = {\mathbf r} - {\mathbf r}_0(t), \label{transkoord}
\end{equation}
where ${\mathbf r}_0(t) =(x_0(t), y_0(t))$ with
\begin{equation} \label{x0y0}
\begin{split}
x_0(t) =& X{\rm e}^{-i\Omega t} + X^*{\rm e}^{i\Omega t},\\
y_0(t) =& Y{\rm e}^{-i\Omega t} + Y^*{\rm e}^{i\Omega t},
\end{split}
\end{equation}
and the amplitudes are
\begin{equation}\label{XY}
X = \frac {e \left(\Omega {\mathcal E}_x - i\omega_c {\mathcal E}_y \right)  }{m^\star\Omega \left(\omega_c^2 - \Omega^2 \right)}, \quad
Y = \frac {e \left(\Omega {\mathcal E}_y + i\omega_c {\mathcal E}_x \right)  }{m^\star\Omega \left(\omega_c^2 - \Omega^2 \right)}.
\end{equation}
Here $\omega_c$ is the cyclotron frequency, which is zero in the absence of an external magnetic field.

Let us consider now an ensemble of electrons in thermal contact with a reservoir. If we apply the above coordinate transformation to each electron and each microscopic element of the reservoir, the many-electron Hamiltonian will also lose the information about the external radiation. This is so because the homogeneous transformation does not change the inter-electron pair interaction. On the other hand, physical properties of the reservoir, which has infinite degrees of freedom, will not be affected. Therefore, in the moving reference frame we can introduce an electron response function coinciding with the one for the same system in the rest reference frame without radiation.

As the next step, let the 2DEG have a single defect, treated as a weak perturbation, with the 2D potential $V_0({\mathbf r})$ in the rest reference frame. We suppose that its internal structure is not developed under the MW field, the defect thus resembling an elementary particle. In the moving reference frame it is only its position that depends on time:
\begin{equation}\label{tildeV0}
\widetilde V_0 ({\mathbf r},t) = S^\dag V_0 ({\mathbf r}) S = V_0 ({\mathbf r} - {\mathbf r}_0(t)).
\end{equation}
For the 2D space Fourier transform of the above we have
\begin{equation}
\widetilde V_0 ({\mathbf q},t) = V_0({\mathbf q}){\mathrm e}^{-i{\mathbf {qr}_0\left(t\right)}}, \label{spaceF}
\end{equation}
where $V_0({\mathbf q})$ is the Fourier transform of the potential of the defect in the rest reference frame and $\mathbf q = (q_x,q_y)$ is a 2D reciprocal coordinate. The time Fourier transform of Eq.~(\ref{spaceF}) yields
\begin{equation}
\widetilde V_0 ({\mathbf q},\omega) = V_0({\mathbf q}) \sum_{n=-\infty}^{+\infty} A_n \left({\mathbf q}\right) \delta \left( n\Omega -\omega\right), \label{timeF}
\end{equation}
where $\omega$ is the frequency domain variable, $\delta (\omega)$ is the Dirac delta function, and for $A_n\left({\mathbf q}\right) = A_n\left( q_x, q_y\right)$ we have
\begin{equation}
A_n\left({\mathbf q}\right) = \frac \Omega {2\pi} \int \limits_{-\pi / \Omega}^{\pi / \Omega } {\mathrm e}^{-i \mathbf q \mathbf r_0 (t) + in \Omega t }\,d t, \label{A_n}
\end{equation}
so that $A^*_n\left({\mathbf q}\right) = A_{-n}\left(-{\mathbf q}\right)$. The potential of the defect screened by the 2DEG is
\begin{equation}
\widetilde V ({\mathbf q},\omega) = \frac {V_0({\mathbf q})} {\varepsilon \left ( q, \omega \right)}\sum_{n=-\infty}^{+\infty} A_n \left({\mathbf q}\right) \delta \left( n\Omega -\omega\right),
\label{tildeV}
\end{equation}
where $\varepsilon \left ( q, \omega \right)$ is the permittivity of the 2DEG without radiation. Here, we do not account for exchange interaction processes, which exist when the defect contains localized electrons. They may be important for low-density 2DEGs, similar to the case of hot electrons in an electron gas. \cite{obmen}

Restoring the direct space and time coordinates for Eq.~(\ref{tildeV}), we have
\begin{equation}
\widetilde V ({\mathbf r},t) = \sum_{n=-\infty}^{+\infty} \int d^2 q \, \frac {V_0({\mathbf q}) A_n \left({\mathbf q}\right) {\mathrm e}^{i\mathbf {qr}-in\Omega t}} {\varepsilon \left ( q, n \Omega \right)}.
\label{tildeVrt}
\end{equation}
In the rest reference frame the screened potential of the defect is
\begin{equation}
V ({\mathbf r},t) = S \widetilde V ({\mathbf r},t) S^\dag = \widetilde V ({\mathbf r + \mathbf r_0 (t)},t).
\end{equation}
Reduced by its bare value, the potential is $\delta V ({\mathbf r},t) = V ({\mathbf r},t) - V_0 ({\mathbf r},t)$:
\begin{widetext}
\begin{equation}
\delta V ({\mathbf r},t) =\sum_{n,\,n'=-\infty}^{+\infty} \int d^2 q \,  V_0({\mathbf q})
\left( \frac 1 {\varepsilon \left( q, n \Omega \right)} -1 \right)
 A_n \left({\mathbf q}\right) A_{n'} \left({- \mathbf q}\right)
{\mathrm e}^{i\mathbf {qr}-i(n+n')\Omega t}.
\label{Vrt}
\end{equation}
\end{widetext}
The remaining part of this paper is devoted to the analysis of Eq.~(\ref{Vrt}). We consider two elementary types of defect: a straight wire breaking the translation symmetry in the $x$ direction or a dot breaking the translation symmetry in both in-plane directions.

\section{Extraction of plasmon field}\label{III_Plasmon_field}

Equation (\ref{Vrt}) resembles the formula for the screened potential of a charged particle moving in plasma. \cite{Neufeld,Echenique} In particular, the radiation harmonics $n\Omega$, $n=\pm 1,\pm 2,\ldots$, play the role of the Doppler frequency shift $\delta \omega = \mathbf {qv}$ in the argument of the response function, where $\mathbf v$ is the velocity of the charged particle. The roots of the equation $\varepsilon \left ( q, n \Omega \right) = 0$ define the wake field or, in other words, plasma excitations, characterized by the wave vector $\Re\mathbf q = \mathbf Q$. The harmonic $n=0$ in Eq.~(\ref{Vrt}) does not relate to the excitation of real plasmons merely contributing to the ``static'' screening, not considered in this paper.

We suppose that the function $V_0({\mathbf q})$ has no poles on the real axis, the oscillatory behavior of the screened potential thus entirely originating due to the plasma excitations. This allows us to calculate the oscillatory part of the potential analytically without numerical evaluation of the integral in Eq.~(\ref{Vrt}), using the plasmon pole expression for the function $\varepsilon^{-1} \left ( q, \omega \right)$ near the plasma resonances, with damping characterized by the relaxation time $\tau >0$:
\begin{equation} \label{epsilon}
\frac 1 {\varepsilon \left( q, \omega \right)} = \frac 1 2
\left( \frac {\omega} {\omega - \omega_p(q)+i/\tau }+\frac {\omega} {\omega + \omega_p(q)+i/\tau }\right).
\end{equation}
Here $\omega_p(q)>0$ represents the spectrum of plasmons, and we suppose that the electron gas is clean enough so that $\Omega \tau \gg 1$.

Actually, the permittivities of the electron gas in the moving and the rest reference frames coincide only in the limit $1/\tau \rightarrow +0$ in the absence of disorder. We have supposed that Eq.~(\ref{epsilon}) still holds for finite $1/\tau$ if in the rest frame the reciprocal permittivity has a similar form but where damping is characterized by the relaxation time $\tau_0 >0$. If the approach is valid, generally $\tau \ne \tau_0$. We can, however, establish conditions in which $\tau \approx \tau_0$ to guarantee the applicability of Eq.~(\ref{epsilon}). For definiteness, we regard disorder due to impurities with the potential $W ( \mathbf r )$, which transforms into the potential $\widetilde W ({\mathbf r},t) = S^\dag W ({\mathbf r}) S$ in the moving reference frame.  For the space-time Fourier transform of the latter, analogously to Eq.~(\ref{timeF}), we have
\begin{equation}
\widetilde W ({\mathbf q},\omega) = W({\mathbf q}) \sum_{n=-\infty}^{+\infty} A_n \left({\mathbf q}\right) \delta \left( n\Omega -\omega\right), \label{stW}
\end{equation}
where $W({\mathbf q})$ is the Fourier transform of $W ( \mathbf r )$. In Eq.~(\ref{stW}), the harmonics with $n \ne 0$  define an effective ``additional'' time-dependent scattering potential, whose contribution to the relaxation time can be taken into account by using the technique developed for electron-phonon scattering. \cite{Ando} Nonetheless, we can estimate its contribution here and determine when it can be neglected. 

Let us introduce the characteristic amplitude $R = \sqrt {\vert Y \vert^2 + \vert X \vert^2}$, where $X$ and $Y$ are given by Eq.~(\ref{XY}). For wave numbers $q \ll 1/R$ we have
\begin{equation}\label{A_0_A_n}
A_0 \left({\mathbf q}\right)= 1 + O  \left( q^2 R^2 \right), \quad A_n(\mathbf q) = O  \left( (q R  )^{\vert n\vert } \right).
\end{equation}
As a result, due to the scalar character of the scattering cross section, which does not depend on the phase of the MW field ${\mathbf E}(t)$, we can assume that
\begin{equation}
\tau = \tau_0 \left( 1 + O  \left( q_{\mathrm {imp}} ^2 R^2 \right) \right),
\end{equation}
where for impurities inside the 2DEG $q_{\mathrm {imp}} \sim 2k_{\mathrm F}$, and $k_{\mathrm F}$ is the electron Fermi wave number, so that we require $2 k_{\mathrm F} R \ll 1$. If the scattering is governed by the presence of charged impurities separated from the 2DEG by a spacer of width $L_{\mathrm {imp}} > 1/2k_{\mathrm F}$,  we have $q_{\mathrm {imp}} \sim 1/L_{\mathrm {imp}}$, and the approximation $\tau \approx \tau_0$ is reached if $R \ll L_{\mathrm {imp}}$.

To estimate the above quantities, as an example, we consider a GaAs/AlGaAs-based 2DEG with concentration $N_s = 3\times 10^{11}$~cm$^{-2}$, radiation with angular frequency $\Omega = 10^{13}$~s$^{-1}$, and the electric field $\mathcal E =30$~V/cm ($\sim 1$~W/cm$^2$ of radiation energy flux). We have $R \sim 1$~\AA\ while $1/k_{\mathrm F} \approx 70 $~\AA\ and typical widths of the spacers are hundreds of \r{a}ngstr\"oms, \cite{Heiblum} which results in the estimate $q_{\mathrm {imp}} ^2 R^2 \lesssim 10^{-3}$. As another example, we can mention that typical values of the MW radiation frequency and the strength of the field used in experiments on MW-induced resistance oscillations are $\Omega \sim (10^{11}$--$10^{12})$~s$^{-1}$ and $\mathcal E \sim 1$~V/cm (see, e.g.,  Ref.~\onlinecite{Zudov2}), so that again $R \lesssim 1$~\AA. These cases demonstrate that the approximation $\tau \approx \tau_0$ can indeed be valid for rather realistic parameters. This guarantees the applicability of Eq.~(\ref{epsilon}), which is still questionable otherwise if $R \gtrsim 1/2k_{\mathrm F}$.

We therefore consider $Q R  \ll 1$. For example, in a GaAs/AlGaAs-based 2DEG with concentration $N_s = 3\times 10^{11}$~cm$^{-2}$, frequency $\Omega = 10^{13}$~s$^{-1}$, and field $\mathcal E =30$~V/cm, the MW field excites 2D plasmons with $Q \approx 2\times 10^5$~cm$^{-1}$ (see below), so that $Q R  \sim 10^{-3}$. Due to Eq.~(\ref{A_0_A_n}), the principal oscillatory part of $\delta V ({\mathbf r},t)$ [see Eq.~(\ref{Vrt})] is enclosed in the potential
\begin{equation}
\begin{split}
V^{(1)} ({\mathbf r},t) = \sum_{n=\pm 1} \int &V_0({\mathbf q})
\left( \frac 1 {\varepsilon \left( q, n \Omega \right)} -1 \right) \\
& \times  A_n \left({\mathbf q}\right)
{\mathrm e}^{i\mathbf {qr}-in\Omega t} \, d^2 q,
\label{Vrt1}
\end{split}
\end{equation}
where we have used $A_0(\mathbf q) \approx 1$. 
From Eq.~(\ref{Vrt}) we can also extract a frozen wave potential, smaller than the principal contribution $V^{(1)} ({\mathbf r},t)$ by the parameter $Q R$:
\begin{equation}
\begin{split}
V^{(0)} ({\mathbf r}) = \sum_{n=\pm 1} \int &
 V_0({\mathbf q}) \left( \frac 1 {\varepsilon \left( q, n \Omega \right)} -1 \right)\\
&\times
\vert A_n \left({\mathbf q}\right)\vert ^2
{\mathrm e}^{i\mathbf {qr}} \, d^2 q.
\label{Vrt0}
\end{split}
\end{equation}
Other harmonics of the excited plasmons can be separated analogously.

\section{Wake from wire defect}\label{IV_Wake_wire}

Let us obtain the excited plasmon field from a straight wire defect placed at $x = 0$, for simplicity having a symmetric space potential, so that its Fourier transform is $V_0({\mathbf q}) = \delta(q_y)V_w(q_x)$ with $V_w(-q_x) = V_w(q_x)$.

The procedure of evaluation of the right-hand sides of Eqs.~(\ref{Vrt1}) and (\ref{Vrt0}) can be illustrated by considering this model integral, corresponding to the terms with $n=1$ in Eqs.~(\ref{Vrt1}) and (\ref{Vrt0}):
\begin{equation}
\mathcal I\left( x\right) = \int \limits_{-\infty}^{+\infty} d q_x\, V_w(q_x)
\left( \frac 1 {\varepsilon \left( q_x, 0, \Omega \right)} -1 \right)
A_1 \left( q_x,0 \right) {\mathrm e}^{iq_xx},
\label{I1}
\end{equation}
where $\varepsilon \left( q_x, 0, \Omega \right) = \varepsilon \left( q, \Omega \right)\vert_{q_y = 0}$.
We will estimate it for the far zone $\vert x \vert > a + 2 \vert X \vert$, where $2a>0$ is the diameter of the wire (see Appendix~\ref{A_wake_far_zone_wire}).

For positive $x>a + 2 \vert X \vert$, the original integration contour may be closed at $\Im q_x \rightarrow +\infty$ by the half-circle path $\vert q_x \vert = \mathrm {const} \rightarrow \infty$,  $\Im q_x \ge 0$. The integral over this path is zero \cite{Sidorov} because the expression in parentheses in Eq.~(\ref{I1}) goes to zero for complex $q_x \rightarrow \infty$. Thus, the right-hand side of Eq.~(\ref{I1}) is entirely defined by the poles of the integrand, the oscillatory behavior originating due to the poles of the function $1/\varepsilon \left( q_x, 0, \Omega \right)$. To find them, we can use Eq.~(\ref{epsilon}). These poles coincide with those of the function $1/(\Omega - \omega_p(q_x,0) + i/\tau)$, where $\omega_p(q_x,0) = \omega_p(q)\vert_{q_y = 0}$, because $\Omega + \omega_p(q_x,0) \ne 0$ due to our adopted choice $\Omega>0$ and $\omega_p(q) >0$. And vice versa, calculating the terms with $n=-1$ in Eqs.~(\ref{Vrt1}) and (\ref{Vrt0}), we deal with the poles of the function $1/( - \Omega + \omega_p(q_x,0) + i/\tau)$ skipping the function $1/(- \Omega - \omega_p(q_x,0) + i/\tau)$.

Continuing with Eq.~(\ref{I1}), we suppose that the poles $q_x = q'_j + i q''_j$ ($j = 1,2,\ldots$) of the integrand are simple, with small imaginary parts: $\vert q''_j \vert \ll \vert q'_j \vert $, where the real parts $q'_j$ are solutions to the equation $\omega_p(q'_j,0) = \Omega$, while
\begin{equation}\label{Im_pole}
q''_j = \frac 1 {\omega '_p(q'_j,0) \tau},
\end{equation}
with $\omega '_p(q_x,0) = d \omega _p(q_x,0) /d q_x$. A general case is briefly discussed in Appendix~\ref{B_arbitrary_spectrum}. The poles are inside the contour of integration if $\omega '_p(q'_j,0) >0$. Accordingly, keeping only the oscillatory contributions from the poles, we have for positive $x>a + 2 \vert X \vert$:
\begin{equation}
\mathcal I \left( x\right) \sim \frac {\pi \Omega} i \sum _{j\, (\omega '_p>0)}
\frac {V_w(q'_j) A_1 ( q'_j,0)} {\omega '_p(q'_j,0)}\, {\mathrm e}^{iq'_j x -q''_j x  }.
\label{I1p}
\end{equation}
Note that $\mathcal I \left( x\right) \rightarrow 0$ as $x \rightarrow +\infty$ due to the exponential decrements $\exp \left( -q''_j x \right)$ with $q''_j >0$ [see Eq.~(\ref{Im_pole})].

Similarly, for negative $x<-a - 2 \vert X \vert$, with the original integration contour closed at $\Im q_x \rightarrow -\infty$, only terms with $\omega '_p(q'_j,0) <0$ should be retained:
\begin{equation}
\mathcal I \left( x\right) \sim i \pi \Omega \sum _ {j\, (\omega '_p <0)} \frac {V_w(q'_j) A_1 ( q'_j,0)} {\omega '_p(q'_j,0)}\, {\mathrm e}^{i q'_jx -q''_j x}.
\label{I1n}
\end{equation}
Again, $\mathcal I \left( x\right) \rightarrow 0$ as $x \rightarrow -\infty$ for $q''_j < 0$ [see Eq.~(\ref{Im_pole})].
\begin{figure}[t]
\includegraphics [width=7cm]{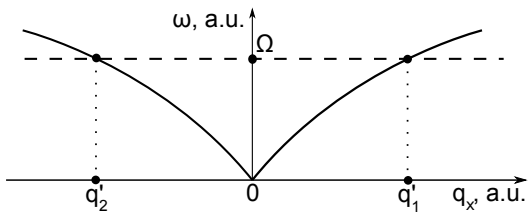}
\caption{Spectrum of 2D plasmons as a function of $q_x$ at $q_y =0$ (full line) and the radiation harmonic $\omega = \Omega$ (dashed line) showing two points of intersection at the wave numbers $q'_1$ and $q'_2 = -q'_1$.
} 
\label{2DP}
\end{figure}

We limit ourselves with the condition of only two plasmon wave numbers contributing to the wake: $q'_1 = -q'_2 = Q >0$. This case is realized for 2D plasmons (see Fig.~\ref{2DP}), having the spectrum $\omega_p(q) = \omega_{2D}(q)$, with $\omega_{2D}(q) = \sqrt{2\pi N_s e^2 q /m^\star\kappa}$, where $\kappa$ is the permittivity of the surrounding medium, generally a function of $q$. \cite{Chaplik} It holds also for magnetoplasmons, at wave numbers much smaller than the reciprocal Larmor radius, with the spectrum $\omega_p(q) = \sqrt {\omega_c^2 +\omega^2_{2D}(q)}$. The same can take place for surface plasmons, depending on the material and the spectral range. \cite{Pitarke} We have plasmons with positive group velocity:
\begin{equation}
\omega '_p(q'_1,0) = - \omega '_p(q'_2,0) = \omega '_p(Q,0)>0.
\end{equation}

For the principal plasmons [see Eq.~(\ref{Vrt1})], we obtain the following expression, valid in the far zone $\vert x \vert > a + 2 \vert X \vert $:
\begin{equation}\label{V1wake}
V^{(1)} (x,t) \sim - \frac {{\mathcal P} _w \mathrm {sgn}\left(x \right) \cos \left( Q\vert x \vert - \Omega t + \psi \right) }
{\exp \left ( \vert x \vert /\omega '_p(Q,0) \tau \right)},
\end{equation}
where the undamped amplitude of the potential of the plasmons is
\begin{equation}
{\mathcal P} _w = \frac {2 \pi V_w(Q) \Omega Q \vert X \vert}{\omega '_p(Q,0)},
\end{equation}
and we have used the approximation $A_1\left( q_x ,0\right) \approx -i q_x X$, while $\psi = \mathrm {arg} X $. The phase velocity of these plasmons is 
\begin{equation}
v_p \left(x,t\right) = \frac {\mathrm {sgn} \left(x \right) \Omega} {Q + \cot \left( Q\vert x \vert - \Omega t + \psi \right)/\omega '_p(Q,0) \tau }.
\end{equation}
They are traveling in the $x$ direction, away from the wire on average [neglecting damping, $v_p \left(x \right) = \mathrm {sgn} \left(x \right) \Omega /Q $]. The case when plasmons with negative group velocity are excited, resulting in plasma waves traveling toward the wire defect, is given in Appendix~\ref{B_arbitrary_spectrum}.

The frozen wave, derived from Eq.~(\ref{Vrt0}), is
\begin{equation}\label{V0wake}
V^{(0)} (x) \sim \frac { {\mathcal P} _w Q \vert X \vert \sin Q \vert x \vert}
{\exp \left ( \vert x \vert /\omega '_p(Q,0) \tau \right)},
\end{equation}
valid in the far zone $\vert x \vert > a + 4 \vert X \vert $; see Appendix~\ref{A_wake_far_zone_wire}. Its amplitude has an extra small parameter $Q \vert X \vert$ as a factor.
 
To determine the structure of the potential ${\mathcal P} _w$, let us assume that the wire defect is small on the scales of the plasmon wavelength, which means that $2aQ \ll 1$. In this case we can neglect the dependence on $Q$ in the Fourier transform $V_w(Q)$:
\begin{equation}
V_w(Q) \approx \frac 1 {2\pi} \int \limits _{-a}^{a} V_w(x) dx \equiv \frac a {\pi} \langle V_w \rangle,
\end{equation}
where we have introduced the average potential $\langle V_w \rangle$ of the wire defect. Considering, for example, 2D plasmons in a medium with a constant permittivity $\kappa$, so that $\omega '_p(q,0) = \omega _p(q,0) /2q$, we have
\begin{equation}
{\mathcal P} _w = 4 \langle V_w \rangle a \vert X \vert Q^2.
\end{equation}
Without magnetic fields, $X \propto 1/\Omega^2$ [see Eq.~(\ref{XY})], so ${\mathcal P} _w  \propto Q$ for a fixed radiation intensity.
This means that without damping it would be more favorable to have excited plasmons with larger wave numbers. Exceptions exist, obviously, for potentials $V_w(Q)$, not small on the scales of the plasmon wavelength, decaying fast enough with $Q$.

The damped amplitude for 2D plasmons is
\begin{equation}\label{damped_ampl}
\bar {\mathcal P} _w = {\mathcal P} _w {\mathrm e}^{- \vert x \vert /\omega '_p(Q,0) \tau }
\propto Q {\mathrm e}^{- 2 \vert x \vert Q / \omega _p(Q,0) \tau}.
\end{equation}
Being a function of the wave number, it reaches a maximum at $Q/\omega _p(Q,0) = \tau / \vert x \vert$. Similar results can be obtained also for 2D magnetoplasmons and surface plasmons, with specific values depending on details of the spectra of the excitations.

Note that the photon-to-plasmon conversion is still more efficient for plasmons with larger wave numbers, even though they might not be well sustained in the system due to damping. Thus, corresponding to the stopping power in the physics of interaction of fast particles with plasma, \cite{Neufeld,Echenique} the dissipative current in a 2D electron system with impurities in quantizing magnetic fields and weak dc electric fields is governed by the excitation of plasmons with the largest wave numbers of the order of $2k_{\mathrm F}$. \cite{Volkov} The concept of the efficiency of the photon-to-plasmon conversion accordingly depends on the goal set: to excite intense persistent plasma waves, or to attain strong absorption of the MW radiation by plasmons, which are excited at the defects. Generally, these goals are reached at different plasmon wave numbers and thus different values of the MW frequency.

There is a plain consequence of the above, important for the problem of the detection of the plasma waves. If a plasmon detector \cite{Muravev} is located at a distance $L$ from a source of plasmons, it is the most efficient for detection of excitations having such wave numbers $Q$ that $1/\omega '_p(Q,0) \sim \tau /L$. For example, for 2D plasmons this condition is $Q/\omega _p(Q,0) = \tau /L $. Plasmons with much larger values of $1/\omega '_p(Q,0)$, which normally correspond to larger $Q$, can not arrive at the detector due to damping, while plasmons with smaller wave numbers are poorly excited.

Finally, let us estimate the distance $L$ for a GaAs/AlGaAs 2D electron system, assuming $\tau = 5 $~ps, which was measured in a system with moderate electron mobility, \cite{Shaner} and using the same parameters as in Sec.~\ref{III_Plasmon_field}: $\Omega = \omega _p(Q,0) = 10^{13}$~s$^{-1}$ and $Q = 2\times 10^5$~cm$^{-1}$. We have $L = 2.5$~$\mu$m or $LQ/2\pi \approx 8$ plasmon wavelengths.

\section{Wake from truncated periodic set of wire defects}\label{V_Wake_wire_set}

For simplicity, in this section we neglect the finite size of the near zone formally considering the limit $Q (a + 4 \vert X \vert ) \rightarrow 0$.

Let us arrange $N$ identical wire defects in a truncated periodic set (grating) with a period $\delta x = \Delta 2\pi / Q$, which is specified by the dimensionless parameter $\Delta$. The values $\Delta = 1,2,\ldots$ provide the plasmon resonance condition. \cite{AllenJr,Theis,Heitmann} The functions
\begin{equation}\label{V10wake_N}
V^{(m)}_N (x,t) = \sum_{n=0}^{N-1} V^{(m)}(x - 2\pi n \Delta / Q,t)
\end{equation}
describe the net potentials of the principal plasma wave ($m=1$) and the frozen plasmon ($m=0$, for which the time argument is superfluous).

\begin{figure}[b]
\includegraphics [width=8.6cm]{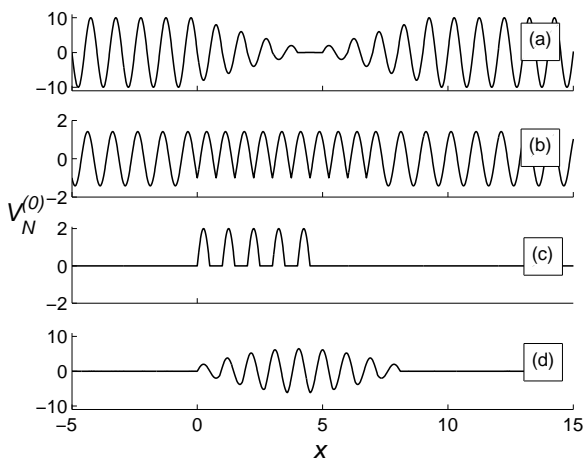}
\caption{Potential (in units of $ {\mathcal P} _w Q \vert X \vert$) of the net frozen plasmon from $N=10$ wires, which form a truncated periodic grating occupying the space $0< x < (N-1) \Delta $ (in units of $2\pi /Q$), for (a) $\Delta = 1$, (b) $\Delta = 3/4$, (c) $\Delta = 1/2$, and (d) $\Delta = 9/10$. Damping is neglected.
} 
\label{qw}
\end{figure}

The sum in Eq.~(\ref{V10wake_N}) can easily be calculated outside the grating region, for $ x \notin \left[ 0, 2\pi(N-1)\Delta/Q \right]$. Neglecting damping for simplicity, we obtain that the amplitude of the net plasmon field is proportional to the parameter
\begin{equation}
\sigma = \Bigg| \frac {\mathrm e ^{ i 2\pi N \Delta} - 1} {\mathrm e ^{ i 2\pi \Delta} -1} \Bigg|.
\end{equation}
At $\Delta = 1,2,\ldots$, we have $\sigma = N$, resulting in $N$-times amplified plasmons outside the grating region: 
\begin{equation}\label{V1wake_N_ampl}
V^{(1)}_N (x,t) \sim - N {\mathcal P} _w \mathrm {sgn}\left(x \right) \cos \left( Q\vert x \vert - \Omega t + \psi \right),
\end{equation}
\begin{equation}\label{V0wake_N_ampl}
V^{(0)}_N (x) \sim N {\mathcal P} _w Q \vert X \vert \sin Q \vert x \vert.
\end{equation}
For non-integer $\Delta$, we have: $\vert \sigma \vert < N$. The destructive interference cases, for which $\sigma = 0$, are obtained for non-integer $\Delta = n/N$ with an integer $n \ne 0$.

To demonstrate the above, in Fig.~\ref{qw} we plot the potential $V^{(0)}_N (x)$ from a grating of $N=10$ wire defects. Presented are the constructive interference case $\Delta = 1$ and several other arrangements including two destructive interference cases at $\Delta = 1/2$ and $\Delta = 9/10$.

\section{Wakes from dot defects}\label{VI_Wake_dots}

\begin{figure}[tb]
\includegraphics [width=5cm]{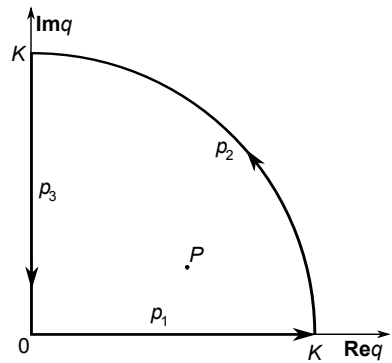}
\caption{Integration contour for Eq.(\ref{I2polar2}) over complex $q$ for the case $\cos \left( \phi - \theta \right) > \lambda >0$. The pole $P$ of the function $\varepsilon^{-1} \left( q, \Omega \right)$ inside the contour defines a plasma wave.}
\label{cont}
\end{figure}

Established for the wire defects, the conclusions on the efficiency of the photon-to-plasmon conversion are valid for a defect of an arbitrary shape. It is the different plasmon wave pattern from a dot defect that makes us pay special attention to this case. In addition, the wake potentials obtained below can play the role of the Green's functions to deal with inclusions of complex shapes.

Let us focus on a dot defect, placed at $x=0,\ y=0$, that has an isotropic potential: $V_0({\mathbf q}) = V_0(q)$. To calculate the integrals in Eqs.~(\ref{Vrt1}) and (\ref{Vrt0}), we analyze the following model integral:
\begin{equation}
\mathcal T\left( \mathbf r \right) = \int d^2 q \, V_0(q) \left( \frac 1 {\varepsilon \left( q, \Omega \right)} -1 \right) A_1 \left( \mathbf q \right) {\mathrm e}^{i\mathbf {qr}}.
\label{I2}
\end{equation}
Using polar coordinates, we transform both the space variables $x = r \cos \theta$, $y = r \sin \theta$, and the integration variables $q_x = q \cos \phi$, $q_y = q \sin \phi$. Being transformed, Eq.~(\ref{I2}) reads:
\begin{equation}
\mathcal T \left( r,\theta \right) =
\int \limits_{0}^{2\pi}  \Phi \, d \phi
\label{I2polar1}
\end{equation}
where $\Phi = \Phi \left( r,\theta,\phi \right)$ is:
\begin{equation}
\begin{split}
\Phi = 
\int \limits_{0}^{+\infty} &
q V_0(q) A_1\left( q \cos \phi, q \sin \phi  \right)\\
&\times \left( \frac 1 {\varepsilon \left( q, \Omega \right)} -1 \right)
{\mathrm e}^{iqr\cos \left( \phi - \theta \right)} \, dq.
\label{I2polar2}
\end{split}
\end{equation}

We will estimate $\mathcal T \left( r,\theta \right)$ for the far zone $r \gg \Lambda_1$ (see Appendix~\ref{C_wake_far_zone_dots}). In Eq.~(\ref{I2polar2}), for $-\pi /2 + \theta + \lambda < \phi < \pi /2 + \theta - \lambda$, the original integration path $p_1$, which is $0\le \Re q < K$ with $\Im q = 0$ and real $K \rightarrow +\infty$ (see Fig.~\ref{cont}) may be supplemented by the quarter-circle path $p_2$, which is $\vert q \vert = K $ and $0<\mathrm {arg} q< \pi /2$, and by the path $p_3$ over the imaginary axis $0\le \Im q < K$ to form a contour.

The integral over the path $p_2$ is zero \cite{Sidorov} if
\begin{equation} \label{limit}
q \left( \frac 1 {\varepsilon \left( q, \omega \right)} - 1\right) \rightarrow 0,
\end{equation}
for  $ q \rightarrow \infty$.
We might validate this condition for a 2DEG with the potential of the inter-electron interaction $V_{ee}(q) \propto 1/q$ by using the fact that the polarizability function $\chi (q) \rightarrow 0$ for $q \rightarrow \infty$. \cite{Stern} Actually, even for finite $\chi (q)$, as obtained in the long-wavelength limit \cite{Stern} or for quantizing magnetic fields, \cite{Chiu} when the left-hand side of Eq.~(\ref{limit}) is finite, the integral over the path $p_2$ is convergent and falls off with distance $r$ without oscillations. The integral over the path $p_3$ also has this property.

As a consequence, the oscillatory behavior of the right-hand side of Eq.~(\ref{I2polar2}) is entirely defined by the poles of the function $1/\varepsilon \left( q, \Omega \right)$ located inside the integration contour. To find them, we can use Eq.~(\ref{epsilon}). The poles coincide with those of the function $1/[\Omega - \omega_p(q) + i/\tau]$ because of the adoption $\Omega>0$ and $\omega_p(q) >0$. Again, calculating the terms with $n=-1$ in Eqs.~(\ref{Vrt1}) and (\ref{Vrt0}), we deal with the poles of the function $1/[ - \Omega + \omega_p(q_x,0) + i/\tau ]$ only.

Repeating the reasoning of Sec.~\ref{IV_Wake_wire}, we suppose that the poles $q_j = q'_j + i q''_j$ ($j = 1,2,\ldots$) are simple and have small imaginary parts: $\vert q''_j \vert \ll \vert q'_j \vert $. For the real parts $q'_j$ we have $\omega_p(q'_j) = \Omega$, while for the imaginary parts $q''_j = 1 / \omega '_p(q'_j) \tau$, where $\omega '_p(q) = d \omega _p(q) /d q$.
The poles are inside the integration contour if $\omega '_p(q'_j) >0$. Keeping only the oscillatory contributions from the poles, we have for $\cos \left( \phi - \theta \right) > \lambda$
\begin{widetext}
\begin{equation}
\Phi_+ = \Phi \left( r,\theta,\phi \right)\Big| _{\cos \left( \phi - \theta \right) > \lambda} \sim \frac {\pi \Omega} i \sum _{j\, (\omega '_p>0)}
\frac {q'_j V_0(q'_j) A_1 \left( q'_j \cos \phi, q'_j \sin \phi  \right) }
{\omega '_p(q'_j)}\, {\mathrm e}^{i\left( q'_j + iq''_j \right)r\cos \left( \phi - \theta \right)}.
\label{I2p}
\end{equation}

Analogously, for $\pi /2 + \theta + \lambda < \phi < 3\pi /2 + \theta - \lambda$, the original integration path in Eq.~(\ref{I2polar2}) should be supplemented by the quarter-circle path $\vert q \vert = \mathrm {const} \rightarrow +\infty$ with $ - \pi /2 <\mathrm {arg} q< 0$ and by the path over the imaginary axis $\Im q < 0$ to form a contour. The plasmon poles are inside the contour of integration if $\omega '_p(q'_j) < 0$. The result is
\begin{equation}
\Phi_- = \Phi \left( r,\theta,\phi \right)\Big| _{\cos \left( \phi - \theta \right) < - \lambda} \sim i \pi \Omega \sum _{j\, (\omega '_p < 0)}
\frac {q'_j V_0(q'_j) A_1 \left( q'_j \cos \phi, q'_j \sin \phi  \right) }
{\omega '_p(q'_j)}\, {\mathrm e}^{i\left( q'_j + iq''_j \right)r\cos \left( \phi - \theta \right)}.
\label{I2m}
\end{equation}

Returning to Eq.~(\ref{I2polar1}), we obtain the following expression:
\begin{equation}
\mathcal T \left( r,\theta \right) \sim 
\int \limits_{- \pi /2 + \theta + \lambda}^{\pi /2 + \theta - \lambda}  \Phi_+ \, d \phi+
\int \limits_{\pi /2 + \theta + \lambda}^{3 \pi /2 + \theta - \lambda}  \Phi_- \, d \phi+
O \left( \Phi \frac {\Lambda_1}{r}\right) 
= \int \limits_{- \pi /2 + \theta}^{\pi /2 + \theta }  \Phi_+ \, d \phi+
\int \limits_{\pi /2 + \theta }^{3 \pi /2 + \theta }  \Phi_- \, d \phi+
O \left( \Phi \frac {\Lambda_1}{r}\right),
\label{I2polar1_parts}
\end{equation}
\end{widetext}
so that we have an asymptotic representation of the function $\Phi \left( r,\theta,\phi \right)$ for the whole domain of the variable $\phi$. The integrals in Eq.~(\ref{I2polar1_parts}) can be evaluated with the help of the approximation
\begin{equation}\label{A_polar}
A_1 \left( q \cos \phi, q \sin \phi  \right) \approx -iq \left( X  \cos \phi + Y \sin \phi\right), 
\end{equation}
and the integrals
\begin{equation}\label{BS_1}
\int\limits_{0}^{\pi} \sin x
\begin{Bmatrix}
\sin \left( y \sin x \right) \\
\cos \left( y \sin x \right)
\end{Bmatrix}
\, dx = \pi 
\begin{Bmatrix}
J_1 \left( y\right)\\
\mathbf H_{-1} \left( y\right)
\end{Bmatrix},
\end{equation}
which are obtained from integral representations of the Bessel function of the first kind $J_m \left( y \right)$ and the Struve function $\mathbf H_m \left( y\right)$: \cite{Bateman}
\begin{equation}
\int\limits_{0}^{\pi}
\begin{Bmatrix}
\sin \left( y \sin x \right) \\
\cos \left( y \sin x \right)
\end{Bmatrix}
\, dx = \pi 
\begin{Bmatrix}
\mathbf H_0 \left( y\right)\\
J_0 \left( y\right)
\end{Bmatrix},
\label{bessel_struve}
\end{equation}
supplemented by derivative identities
\begin{equation}\label{derivative}
\frac d {dy} \left( y^m
\begin{Bmatrix}
J_m \left( y\right) \\
\mathbf H_m \left( y\right)
\end{Bmatrix} \right) = 
y^m
\begin{Bmatrix}
J_{m-1} \left( y\right) \\
\mathbf H_{m-1} \left( y\right)
\end{Bmatrix}
\end{equation}
and the property $J_{-m} (y) = (-1)^m J_m(y)$ (the latter is valid for integer $m$). To evaluate the potential of the frozen wave, given by Eq.~(\ref{Vrt0}), we also use the following result deduced from Eqs.~(\ref{bessel_struve}) and (\ref{derivative}):
\begin{equation}\label{BS_2}
\int\limits_{0}^{\pi} \sin^2 x
\begin{Bmatrix}
\sin \left( y \sin x \right) \\
\cos \left( y \sin x \right)
\end{Bmatrix}
\, dx = \pi 
\begin{Bmatrix}
- \frac {\mathbf H_{-1} \left( y\right)} y - \mathbf H_{-2} \left( y\right)\\
 \frac {J_1 \left( y\right)} y - J_2 \left( y\right)
\end{Bmatrix}.
\end{equation}

Having established the integration procedure, we merely present the result for the right-hand sides of Eqs.~(\ref{Vrt1}) and (\ref{Vrt0}). We suppose that $Q>0$ is a single solution to the equation $\omega_p(Q) = \Omega$ so that $\omega '_p(Q) > 0$ (see Fig.~\ref{2DP}). For circular polarizations of the MWs, ${\mathcal E}_y = \pm i {\mathcal E}_x$, the principal wave is a circular plasmon corotating with the electric field vector of the radiation [counterrotating if $\omega '_p(Q) < 0$]:
\begin{equation}
V^{(1)} (r,\theta,t) \sim - \frac {\mathcal P_d} 2 \Bigl( \mathrm e ^{\pm i \theta - i \Omega t + i \psi} F_1(r) + \mathrm {c.c.} \Bigr),
\label{Vrt1d}
\end{equation}
where the undamped amplitude of the wave is
\begin{equation}
\mathcal P_d =  \frac {2 \pi^2 V_0(Q) Q^2 \Omega \vert X \vert }{\omega '_p(Q)}.
\end{equation}
$\mathrm {c.c.}$ stands for the complex conjugate of the preceding term in parentheses of Eq.~(\ref{Vrt1d}), and, as a reminder, $\psi = \mathrm {arg} X $. The complex radial function $F_1(r)$ is
\begin{widetext}
\begin{equation}\label{radial}
F_1 \left( r\right) = i J_1 \left(  Qr + \frac {ir} {\omega '_p(Q)\tau} \right)
+ \mathbf H_{-1} \left(  Qr + \frac {ir} {\omega '_p(Q)\tau} \right)
 + O \left( \frac {\Lambda_1}{r}\right).
\end{equation}
\end{widetext}

Let us inspect the asymptotics of the Bessel and Struve functions \cite{Bateman}
\begin{equation}\label{Bessel_asimpt}
J_m \left(\xi \right) = \sqrt {\frac 2 {\pi \xi}} \cos \left( \xi - \frac {m \pi} 2 - \frac \pi 4 \right)  + O \left( \xi ^ {-1}\right),
\end{equation}
\begin{equation} \label{Struve_asimpt}
\mathbf H_m \left(\xi \right) = \sqrt {\frac 2 {\pi \xi}} \sin \left( \xi - \frac { m \pi} 2 - \frac \pi 4 \right) + O \left( \xi ^ {m-1}\right),
\end{equation}
valid for $ \xi \to \infty $ [below we use $\mathbf H_m\left(\xi \right)$ only for $m \le 0$, so that the form of Eq.~(\ref{Struve_asimpt}) is adequate for our goals]. Using Eqs.~(\ref{Bessel_asimpt}) and (\ref{Struve_asimpt}) in Eq.~(\ref{radial}), we obtain the asymptotics for the function $F_1 \left( r\right)$:
\begin{equation}
F_1 \left( r\right) \sim \sqrt {\frac 2 {\pi Q r }} \ \mathrm e ^{i Qr - r/\omega '_p(Q)\tau - i \pi /4},
\end{equation}
as $Qr \rightarrow \infty$. Note that we have used $Q \gg 1 / \omega '_p(Q) \tau$ under the square root in the expression above.

For the frozen plasmon induced by MWs with circular polarization we have the following isotropic potential:
\begin{equation}
V^{(0)} (r) \sim - \frac {\mathcal P_d Q \vert X \vert} 2 \Bigl( F_0(r) + \mathrm {c.c.} \Bigr),
\label{Vrt0d}
\end{equation}
where the complex radial function $F_0(r)$ is
\begin{widetext}
\begin{equation}\label{radial0}
F_0 \left( r\right) = i J_0 \left(  Qr + \frac {ir} {\omega '_p(Q)\tau} \right)
- \mathbf H_0 \left(  Qr + \frac {ir} {\omega '_p(Q)\tau} \right)
 + O \left( \frac {\Lambda_0}{r}\right),
\end{equation}
\end{widetext}
valid for the far zone $r \gg \Lambda_0$ (see Appendix~\ref{C_wake_far_zone_dots}). The function $F_0 \left( r\right)$ has the asymptotics, similar to $F_1 \left( r\right)$
\begin{equation} \label{F0_as}
F_0 \left( r\right) \sim i \sqrt {\frac 2 {\pi Q r }} \ \mathrm e ^{i Qr - r/\omega '_p(Q)\tau - i \pi /4}
\end{equation}
for $Qr \rightarrow \infty$. When compared to the principal plasmon, the amplitude of the frozen wake has an extra small parameter $Q \vert X \vert$ as a factor, similar to the wire defect case [see Eq.~(\ref{V0wake}].

Let us obtain the potentials $V^{(1)} (r,\theta,t)$ and $V^{(0)} (r,\theta)$ for the case of a linear polarization of MWs, for example, for $X \ne 0$ and $Y = 0$ (without magnetic field this is equivalent to ${\mathcal E}_y =0$). We obtain the principal plasmon having the character of a standing wave:
\begin{equation}
V^{(1)} (r,\theta,t) \sim - \frac {\mathcal P_d} 2  \Bigl( \mathrm e ^{- i \Omega t + i \psi} F_1(r) + \mathrm {c.c.} \Bigr) \cos \theta,
\label{Vrt1d_stand}
\end{equation}
with the maximum amplitude along the $x$ axis, $\theta =0$. For the frozen plasmon we have the following result:
\begin{equation}
V^{(0)} (r, \theta) \sim - \frac {\mathcal P_d Q \vert X \vert} 2 \Bigl( 
F_0(r) \sin^2\theta   + F_2(r) \cos 2\theta + \mathrm {c.c.} \Bigr),
\label{Vrt0d_stand}
\end{equation}
where the complex radial function $F_2(r)$ is
\begin{widetext}
\begin{equation}\label{radial2}
F_2 \left( r\right) = i \frac {J_1 \left(  Qr + \frac {ir} {\omega '_p(Q)\tau} \right)}{Qr} -
i J_2 \left(  Qr + \frac {ir} {\omega '_p(Q)\tau} \right)
+ \frac {\mathbf H_{-1} \left(  Qr + \frac {ir} {\omega '_p(Q)\tau} \right)}{Qr}
+ \mathbf H_{-2} \left(  Qr + \frac {ir} {\omega '_p(Q)\tau} \right)
 + O \left( \frac {\Lambda_0}{r}\right),
\end{equation}
\end{widetext}
which has the asymptotics $F_2 \left( r\right) \sim F_0 \left( r\right)$, so that in the limit $Qr \rightarrow \infty$ we have the following result for the frozen wave: 
\begin{equation}
V^{(0)} (r, \theta) \sim - \frac {\mathcal P_d Q \vert X \vert} 2 \Bigl( F_0(r) + \mathrm {c.c.} \Bigr) \cos^2 \theta ,
\label{Vrt0d_stand_as}
\end{equation}
where for the function $F_0(r)$ one should use Eq.~(\ref{F0_as}).
\begin{figure}[bh]
\includegraphics [width=8.6cm]{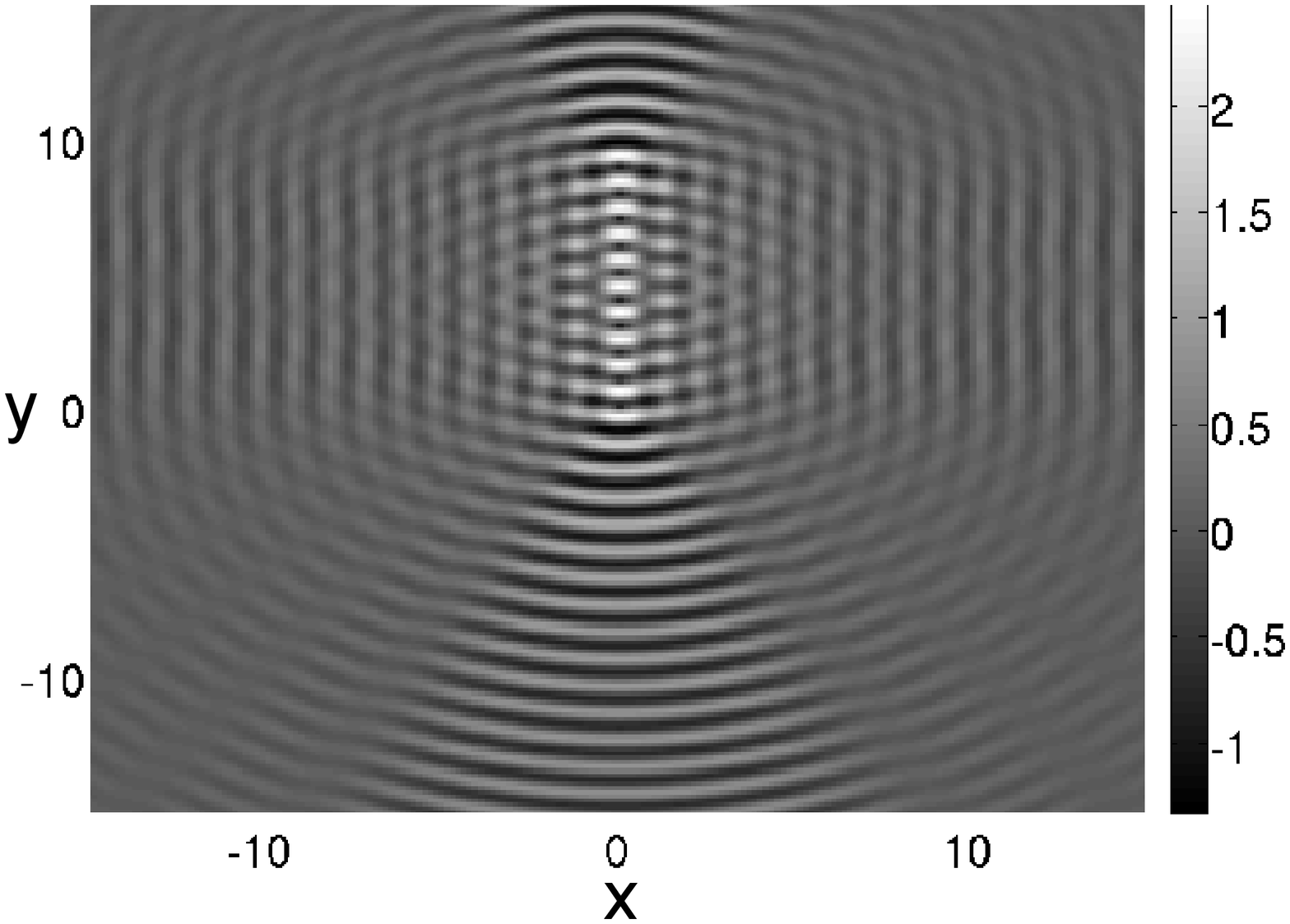}
\caption{
Potential (in units of $ {\mathcal P} _d Q \vert X \vert$) of the net frozen plasmon from $N=10$ dots, which form a truncated periodic set occupying the space $x=0$, $0< y < (N-1)$ (in units of $2\pi /Q$). Constructive interference. Circular polarization. Damping is neglected.
}
\label{dots_cc}
\end{figure}

\begin{figure}[bh]
\includegraphics [width=8.6cm]{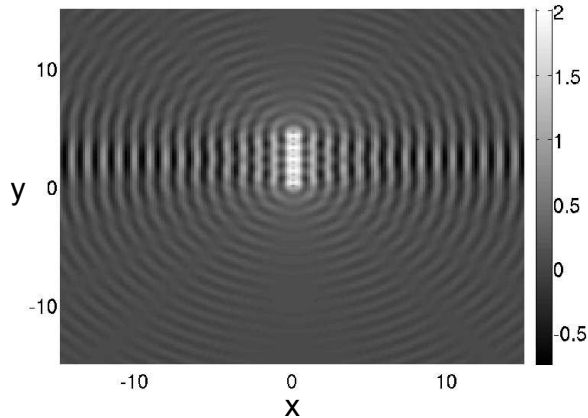}
\caption{Potential (in units of $ {\mathcal P} _d Q \vert X \vert$) of the net frozen plasmon from $N=10$ dots, which form a truncated periodic set occupying the space $x=0$, $0< y < (N-1)/2$ (in units of $2\pi /Q$). Destructive interference. Circular polarization. Damping is neglected.}
\label{dots_dc}
\end{figure}

Finally, to amplify the plasmon wave with a set of dot defects, we should arrange them in a truncated periodic set that again has the fundamental lattice period $2\pi /Q$. Due to the fact that even without damping the plasma wave from a dot defect decays with distance as $\propto 1/\sqrt{Qr}$, the interference pattern has relatively low contrast. To demonstrate this, in Figs.~\ref{dots_cc} and \ref{dots_dc} we plot the function
\begin{equation}\label{V10wake_N_D}
V^{(0)}_N (x,y) = \sum_{n=0}^{N-1} V^{(0)} \left( x,
y-2\pi n \Delta/ Q \right),
\end{equation}
which represents the frozen plasmon wave produced by a linear chain of $N$ dots separated by the distance $2\pi \Delta /Q$, for two values $\Delta =1$ and $\Delta =1/2$ for circular polarization of the radiation.

\section{Role of retardation and polarization of defects}\label{VII_Role_missing}

We have neglected the retardation of the electromagnetic interaction, focusing on plasma excitations with the wave numbers $Q \gg k$, where $k$ is the wave number of the MW radiation. This has provided us with the approximation that the spectra of plasmons in the rest and the moving reference frames coincide with each other (because Poisson's equation is invariant with respect to a uniform coordinate transformation). The plasmons thus obtained do not experience the radiative decay, \cite{ChiuLeeQuinn} the effect being recovered by the inclusion of the retardation. \cite{Satou} In a clean 2DEG, this makes the excitation of plasmons with larger wave numbers even more efficient because they are less prone to decay.

The retardation can be implemented in the developed method by following the line of the theory of interaction of relativistic particles with matter, \cite{Landavshitz} although it is complicated by the necessity to allow also for the spatial dispersion of the conductivity tensor. \cite{Chiu} Note that Maxwell's equations change in the moving reference frame due to the non-Lorentz-invariant transformation realized by the operator $S$. The excited plasmons can still be obtained analytically for wave numbers $Q\ll 1/R$. On the other hand, the existing approaches, \cite{Sondergaard,Rotenberg,Satou} which are devoid of the relativistic limitations, would require extensive numerical simulations if they were modified to take into account collective excitations with large wave numbers. Such are in particular plasmons with negative group velocity, which produce backward propagating plasmon waves, readily obtained within the wake-field method as long as their spectrum $\omega_p(q)$ and the relaxation time $\tau$ are known.

A more delicate issue is the radiation-induced polarization of the defects, leading to the scattering of the incoming MWs. We have neglected it, regarding the defects as elementary particles lacking an internal structure. Strictly speaking, this approximation is valid only for defects that are represented by quantum mechanical objects with the excited states lying far above the energy of the radiation, and in weak radiation fields so that the dipole moment of the defects multiplied by the electric field is small as compared to the energy of the excited states. This is not a typical situation. On the contrary, the defects are usually classical objects: ripplons, \cite{Grimes} lattices of metal scatterers and gratings, \cite{AllenJr,Allen,Barnes} or surface acoustic waves. \cite{Kukushkin2}

To make allowance for the general situation, we should first determine the potential $V^{(\mathrm {scat})}_0 \left( \mathbf r , t \right)$ of the scattered wave and carry on with its sum with the potential of the bare defect forming the total unscreened potential $V^{(\mathrm {tot})}_0$:
\begin{equation}
V^{(\mathrm {tot})}_0 \left( \mathbf r , t \right) = V_0 \left( \mathbf r \right)
+V^{(\mathrm {scat})}_0 \left( \mathbf r , t \right).
\end{equation}
The scattered wave contains frequency harmonics $\Omega_\mathrm s$, discrete for brevity's sake, which may include those not present in the external field $\mathbf E (t)$:
\begin{equation}
V^{(\mathrm {scat})}_0 \left( \mathbf r , t \right) = \sum_{\Omega_\mathrm s} V^{(\mathrm {scat})}_0 \left( \mathbf r , \Omega_\mathrm s \right) \mathrm e ^{i \Omega_\mathrm s t}.
\end{equation}
It suffices to mention a two-level system whose dipole moment oscillates not only at the frequencies $\Omega_\mathrm s = \pm \Omega$ of the external field but also at the frequencies $\Omega_\mathrm s = \pm \Omega \pm \Omega_{\mathrm R}$, where $\Omega_{\mathrm R}$ is the generalized Rabi frequency. \cite{Boyd}

In the moving reference frame we naturally have
\begin{equation}\label{tildeVtot}
\begin{split}
\widetilde V^{(\mathrm {tot})}_0 ({\mathbf r},t) &= S^\dag V^{(\mathrm {tot})}_0 ({\mathbf r}) S\\
&= V_0 ({\mathbf r} - {\mathbf r}_0(t)) + V^{(\mathrm {scat})}_0 \left( \mathbf r - {\mathbf r}_0(t), t \right).
\end{split}
\end{equation}
This potential is then screened by the 2DEG and processed in the straightforward way developed above.

For the principal plasma wave, in the linear regime, we can neglect the weak MW-induced polarization of the 2DEG in the scattered wave potential. This is equivalent to the following approximation:
\begin{equation}\label{tildeVtot_approx}
\widetilde V^{(\mathrm {tot})}_0 ({\mathbf r},t)
\approx V_0 ({\mathbf r} - {\mathbf r}_0(t)) + V^{(\mathrm {scat})}_0 \left( \mathbf r , t \right).
\end{equation}
In this approximation, the excited plasma wave consists of two contributions: the one originating due to the MW-induced polarization of the electron gas and scattering of these polarization waves at bare defects, considered in detail in this paper, and the contribution that is the MW-induced polarization potential of the defects screened by the MW-unpolarized electron gas. The latter is sometimes treated as the only source of the excited plasmons, \cite{Novotny} which is reasonable if the defects are not charged and located far enough from the electron gas so that they weakly influence in particular the dc conductivity of the electron gas. These two excitation mechanisms cannot be separated in the local permittivity- or conductivity-based approaches. \cite{Sondergaard,Rotenberg,Satou}

To complete the analysis, let us compare the strengths of these two contributions for a defect given by a metal sphere with Coulomb charge $Z$, radius $R_0$, and center with the coordinates $(x=0,\,y=0,\,z=d)$, isolated from a 2DEG at $z=0$, so that $d > R_0$. The system is placed in a medium with permittivity $\kappa$. The length $d$ is small as compared to the wavelength of the MW radiation in the medium $ 2\pi c /\Omega \sqrt \kappa $, where $c$ is the speed of light in vacuum.  Under these conditions, the 2D potential of the defect is
\begin{equation}\label{kulon}
 V_0 (q) = \frac Z {2\pi \kappa q} \, \mathrm e ^{-qd }.
\end{equation}
The 3D polarization field in the conditions of small distances is \cite{Landavshitz2} 
\begin{equation}
V^{(\mathrm {scat\, 3D})}_0 (\mathbf r, z,t) \approx \frac {R^3_0 \mathbf r \cdot \mathbf E (t)} {(r^2 +(z-d)^2)^{3/2}}
\end{equation}
[we have previously defined $\mathbf r$ and $\mathbf E(t)$ as 2D vectors], which corresponds to the polarization of a sphere in a static homogeneous electric field. \cite{Landavshitz} The spatial Fourier transform of the 2D scattered wave potential is thus
\begin{equation}
V^{(\mathrm {scat})}_0 (\mathbf q,t)=\frac {i R^3_0}{2\pi q} \, \mathrm e ^{-qd} \mathbf q \cdot \mathbf E (t) .
\end{equation}
This potential screened by the 2DEG and reduced by its bare value has the following form in the space-time coordinates:
\begin{equation}
\begin{split}
V^{(\mathrm {scat})} ({\mathbf r},t) = &\sum_{n=\pm 1} \int \frac {i R^3_0}{2\pi q}\,\mathrm e ^{-qd}
\left( \frac 1 {\varepsilon \left( q, n \Omega \right)} -1 \right) \\
& \times  \left( q_x {\mathcal E}^{(n)}_x + q_y {\mathcal E}^{(n)}_y\right) 
{\mathrm e}^{i\mathbf {qr}-in\Omega t} \, d^2 q,
\label{Vrt1scat}
\end{split}
\end{equation}
where ${\mathcal E}^{(1)}_j = {\mathcal E}_j$ and ${\mathcal E}^{(-1)}_j = {\mathcal E}^*_j$.

The structure of Eq.~(\ref{Vrt1scat}) is identical to that of Eq.~(\ref{Vrt1}) in which the potential is expressed by Eq.~(\ref{kulon}). Moreover, without magnetic field, $X \propto {\mathcal E}_x$ and $Y \propto {\mathcal E}_y$, so that the two excitation mechanisms produce additive contributions, with the relative sign being determined by the sign of the charge $Z$ [see Eqs.~(\ref{XY}) and (\ref{A_polar})]. In finite magnetic fields these waves interfere. Let us finally compare their relative strength, at $\omega_c =0$ for definiteness. We introduce the electric potential of the sphere $\Pi = Z/\kappa R_0$ and the wavelength of the MW radiation in vacuum, $ l = 2\pi c /\Omega $, and analyze 
the following dimensionless parameter:
\begin{equation}
{\cal R} = \frac {e\Pi} {4\pi^2 m^\star c^2} \left( \frac l {R_0}\right)^2.  
\end{equation}
If ${\cal R} \gg 1$, the polarization of the defect can be neglected even though it is a classical object. To check whether this is reasonably achievable in a GaAs/AlGaAs-based 2DEG, we use the following benchmark data: decent $e\Pi = 1$~eV while $4\pi^2 m^\star c^2 \approx 1.4$~MeV. So we require that $R_0 \ll l \times 10^{-3}$, and if $R_0 \sim 1\ \mu$m, the ``elementary particles'' mechanism dominates for frequencies $\Omega/2\pi \ll 0.3 $~THz. This figure is somewhat below the range of interest for actual applications, \cite{Allen} yet there is a definite reserve in increase of the potential $\Pi$ and decrease of the size of the defects $R_0$.

\section{Conclusions}\label{VIII_conclusion}

We have developed an efficient method to find plasmons excited by MWs with a homogeneous field in a 2DEG with parabolic spectrum having defects, treated as weak scatterers. The method uses the concept of the wake created by a charged particle moving in plasma. The expressions that we have derived for excited plasmons explicitly address the efficiency of the photon-to-plasmon conversion and the type of the resulting excitations.

We have shown that the photon-to-plasmon conversion is more efficient for plasmons with larger wave numbers, even though they might not be well sustained in the system due to damping. Intense persistent plasma waves are created at optimal wave numbers that depend on the spectrum of plasmons and the distance that the waves are required to travel without being substantially attenuated.

Concerning the type of excited plasmons, they can have traveling or standing wave components, depending on the geometry of the defects and the polarization of the radiation. Thus, a single straight wire defect can produce traverse plasmons traveling away from the wire if their group velocity is positive [see Eq.~(\ref{V1wake})] and toward the wire if it is negative [see Appendix~\ref{B_arbitrary_spectrum}]. A single dot defect can excite circular plasmons [see Eq.~(\ref{Vrt1d})] when the MW radiation has circular polarization, or standing plasmon waves for linearly polarized MWs [see Eq.~(\ref{Vrt1d_stand})]. The zeroth harmonic of the excited plasmons, which is a frozen charge density wave, has also been obtained [see Eq.~(\ref{V0wake}) for a wire defect and Eqs.~(\ref{Vrt0d}) and (\ref{Vrt0d_stand}) for a dot defect]. Defects arranged in truncated periodic sets can create amplified interference patterns, which can be found easily [see Eqs.~(\ref{V1wake_N_ampl}) and (\ref{V0wake_N_ampl}) and Fig.~\ref{qw} for wire defects, and Fig.~\ref{dots_cc} for dot defects].

\section*{Acknowledgments}

We are grateful to the referee for comments, addressing which led to an improvement of the content. This work was supported by the NSERC and CRC Program, Canada.

\appendix

\section{Definition of wake far zone for wire defect}\label{A_wake_far_zone_wire}

If we specify the boundaries of the wire as $\vert x \vert = a$, for the Fourier transform $V_w(q_x)$ of its potential $V_w(x)$ we can use the following expression:
\begin{equation}\label{Vwire}
V_w(q_x) = \frac 1 {2\pi} \int \limits _{-a}^{a} V_w(x) \mathrm e ^{-iq_x x} dx.
\end{equation}
Let us rewrite Eq.~({\ref{I1}) explicitly using Eqs.~(\ref{A_n}) and (\ref{Vwire}) for $A_1 \left( q_x,0 \right)$ and $V_w(q_x)$:
\begin{widetext}
\begin{equation}
\mathcal I \left( x\right) = \frac {\Omega} {\left( 2\pi\right)^2} \int\limits_{-\pi/\Omega}^{\pi/\Omega} dt \, \mathrm e ^{i\Omega t}
\int\limits_{-a}^{a} dx' \, V_w \left( x' \right)
\int\limits_{-\infty}^{+\infty} dq_x \, \left( \frac 1 {\varepsilon \left( q_x, 0, \Omega \right)} -1 \right) \,
\mathrm e ^{iq_x \left[ x - x' - x_0\left( t \right) \right]}.
\label{I1explicit}
\end{equation}
\end{widetext}
The expression in square brackets in Eq.~({\ref{I1explicit}) is positive if $x > a + 2 \vert X \vert$ and negative if $x < -a - 2 \vert X \vert$ for the entire domain spanned by the integration variables: $ -a\le x' \le a $ and $-\pi/\Omega \le t \le \pi/\Omega$. Thus, we define the far zone, $\vert x\vert > a + 2 \vert X \vert$, as the region where structural details of the functions $A_1 \left( q_x,0 \right)$ and $V_w(q_x)$ do not influence the shape of the wake.
The value obtained for the position of the far zone boundary is applicable to the principal plasmon given by Eq.~(\ref{Vrt1}). The right-hand side of Eq.~(\ref{Vrt0}) has an extra factor $A_n$ under the integral, which results in an extension of the far zone to $\vert x\vert > a + 4 \vert X \vert$ for the frozen wave $V^{(0)} ( x )$.

\section{Excitation of plasmons with arbitrary spectrum}\label{B_arbitrary_spectrum}

\begin{figure}[tbh]
\includegraphics [width=6cm]{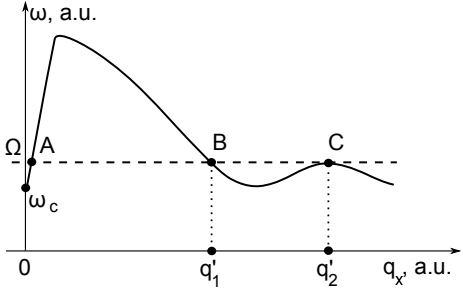}
\caption{Spectrum of the principal 2D magnetoplasmon mode in weak magnetic fields as a function of $q_x$ at $q_y =0$ (full line) and the radiation harmonic $\omega = \Omega$ (dashed line). Negative $q_x$ are not shown. Contributions of the plasmons with the wave numbers $q_x = q'_1$ (the intersection point $B$) and $q_x = q'_2$ (point $C$) are discussed in Appendix~B. Interaction of photons with plasmons at smaller wave numbers (point $A$) is detailed in the main text.
}
\label{MPLQ}
\end{figure}
Let us briefly discuss different situations that may arise while considering collective excitations with arbitrary spectra. As an example, in Fig.~\ref{MPLQ} we present the spectrum of the principal 2D magnetoplasmon mode in a weak magnetic field (several Landau levels are filled), obtained in the random phase approximation. \cite{Chiu} Negative $q_x$ are not shown, but inferred. For $q_x > 0$ with $q_y =0$, the spectrum and the radiation harmonic $\omega = \Omega$ have three common points.  The case corresponding to the point $A$ is discussed in the main text. The point $B$ (at $q_x = q'_1$) is treated analogously to the point $A$, but here the radiation excites plasmons with negative group velocity $\omega '_p (q'_1,0) <0$ at positive $q_x = q'_1$ with a complementary $\omega '_p (-q'_1,0) >0$ at negative $q_x = -q'_1$ (not shown). This generates a component of the principal plasmon wave, analogous to the one given by Eq.~(\ref{V1wake}), traveling {\em toward} the wire defect with the phase velocity $v_p \left(x \right) =  - \mathrm {sgn} \left(x \right) \Omega /q'_1$ (neglecting damping).

The point $C$ is a tangency point, for which $\omega '_p (q'_2,0) =0$. The approach used in the main text is not valid now. Actually, considering the model integral of Eq.~(\ref{I1}), we must retreat to find the roots of the equation
\begin{equation}\label{b1}
\omega _p (q_x,0) = \Omega +i/\tau. 
\end{equation}
For weak damping, we can use the next term in the Taylor series expansion around the point $q '_2$:
\begin{equation}\label{b2}
\omega _p (q_x,0) \approx \omega _p (q '_2,0) + \omega ' _p (q '_2,0) \delta q + \omega '' _p (q '_2,0) \frac {\delta q^2} 2, 
\end{equation}
where $\delta q = q_x - q '_2$ and $\omega _p (q '_2,0) =\Omega$. The expansion given by Eq.~(\ref{b2}) can be used when the third term in the right-hand side is not small as compared to the second one, in particular, at the tangency point $C$ where $\omega '_p (q'_2,0) =0$. In this case from Eqs.~(\ref{b1}) and (\ref{b2}) we have $\delta q ^2 \approx 2i /\omega '' _p (q'_2, 0) \tau$, which gives two complex values of $\delta q$. The poles of the integrand in Eq.~(\ref{I1}) remain simple and located at $q_x = q'_2 + \delta q$. Only one of them falls inside the integration contour for positive $x$, the other contributing for negative $x$.  This gives two extra plasma waves (totaling four with points $A$ and $B$) propagating in opposite directions: the one traveling away from the wire has the wave number $Q_+$, while the wave traveling toward the wire has the wave number $Q_-$, where
\begin{equation}
Q_\pm = q'_2 \pm \mathrm {sgn} \left( \omega '' _p (q'_2, 0)\right) \sqrt { \frac 1 {\vert \omega '' _p (q'_2, 0) \vert \tau}}.
\end{equation}
Both waves are characterized by the exponential decrement $\exp \left( -\vert x \vert / \sqrt {\vert \omega '' _p (q'_2, 0) \vert \tau}\right)$.

\section{Wake far zone for dot defect}\label{C_wake_far_zone_dots}

If the dot defect occupies the area $\pi D^2$, where $D>0$ is the nominal radius of the defect, for the Fourier transform $V_0(q)$ of its potential $V_0(r) \equiv V_0(x,y)$ we can use the following expression:
\begin{equation}\label{Vdot}
V_0(q) = \iint \limits _{r'<D} \frac {V_0(x',y')}{\left( 2\pi \right)^2} \, \mathrm e ^{-iq\left( x'\cos \phi + y'\sin \phi \right)} \, dx' dy',
\end{equation}
where we have used polar coordinates for the reciprocal space, as in Eq.~({\ref{I2polar2}), and have kept the superfluous dependence on the angular coordinate $\phi$ in the integrand.
Let us rearrange Eq.~({\ref{I2polar2}) explicitly using Eqs.~(\ref{A_n}) and (\ref{Vdot}) for $A_1 \left( q \cos \phi, q \sin \phi  \right)$ and $V_0(q)$:
\begin{widetext}
\begin{equation}
\Phi =  \frac {\Omega}{\left( 2\pi \right)^3}
\int\limits_{-\pi/\Omega}^{\pi/\Omega} dt \, \mathrm e ^{i\Omega t}
\iint \limits _{r'<D} dx' dy' \, V_0(x',y') \int\limits_{0}^{+\infty} dq \, q
\, \left( \frac 1 {\varepsilon \left( q, \Omega \right)} -1 \right) \,
\mathrm e ^{iq \hat \rho },
\label{I2polar_expl}
\end{equation}
\end{widetext}
where the scalar quantity $\hat \rho$, which is a function of the variables $\left(x', y', t \right)$ and the parameters $\left( r, \theta , \phi \right)$, is
\begin{equation}
\hat \rho = r\cos \left( \phi - \theta \right) -  \left[ x' + x_0\left( t \right) \right] \cos \phi  - \left[ y' + y_0\left( t \right) \right] \sin \phi .
\end{equation}
Our goal is to find conditions when this operator is positive and negative definite. Such conditions must depend neither on the variable $t$, which characterizes the MW-related coordinates $x_0\left( t \right)$ and $y_0\left( t \right)$, nor on the variables $x'$ and $y'$, which specify the structure of the defect. Then, as in the case of a wire defect, the shape of the wake will not depend on structural details of the functions $A_1 \left( \mathbf q \right)$ and $V_0(q)$.

In contrast to the wire defect case, the boundary of the far zone can not be rigorously defined. Let us introduce the following quantity:
\begin{equation}
\Lambda_1 = \sqrt {\left( D + 2 \vert X \vert \right)^2 + \left( D + 2 \vert Y \vert \right)^2}.
\end{equation}
Then, for $r \gg \Lambda_1$ we have
\begin{equation}
\begin{cases}
\hat \rho > 0, & -\frac \pi 2 + \lambda < \phi - \theta < \frac \pi 2 - \lambda,\\
\hat \rho < 0, & \frac \pi 2 + \lambda < \phi - \theta < \frac {3 \pi} 2 - \lambda,
\end{cases}
\end{equation}
where $\lambda \approx  \Lambda_1 / r \rightarrow 0$ as $r \rightarrow \infty$.

The expression that we obtain in the main text for the plasma wave created by a dot defect is valid in the asymptotic limit $r / \Lambda_1 \rightarrow \infty$.

For the frozen plasmon, the right-hand side of Eq.~(\ref{Vrt0}) has an extra factor $A_n$ under the integral, which makes us introduce
\begin{equation}
\Lambda_0 = \sqrt {\left( D + 4 \vert X \vert \right)^2 + \left( D + 4 \vert Y \vert \right)^2},
\end{equation}
and deal with $V^{(0)} (r)$ in the asymptotic limit $r / \Lambda_0 \rightarrow \infty$.


\begin{thebibliography}{99}

\bibitem{Grimes} C.C.~Grimes and G.~Adams, Phys.\ Rev.\ Lett. {\bf 36}, 145 (1976).

\bibitem{AllenJr} S.J.~Allen, Jr., D.C.~Tsui, and R.A.~Logan, Phys.\ Rev.\ Lett. {\bf 38}, 980 (1977).

\bibitem{Theis} T.N.~Theis, Surf.\ Sci. {\bf 98}, 515 (1980).

\bibitem{Heitmann} D.~Heitmann, Surf.\ Sci. {\bf 170}, 332 (1986). 

\bibitem{Kukushkin1} I.V.~Kukushkin, J.H.~Smet, S.A.~Mikhailov, D.V.~Kulakovskii, K.~von~Klitzing, and W.~Wegscheider, Phys.\ Rev.\ Lett. {\bf 90}, 156801 (2003).

\bibitem{Kukushkin2} I.V.~Kukushkin, J.H.~Smet, V.W.~Scarola, V.~Umansky, and K.~von~Klitzing, Science {\bf 324}, 1044 (2009).

\bibitem{Dyakonov} M.~Dyakonov and M.~Shur,  IEEE Trans.\ Electron. Devices {\bf 43}, 380 (1996).

\bibitem{Allen} X.G.~Peralta, S.J.~Allen, M.C.~Wanke, N.E.~Harff, J.A.~Si\-m\-mons, M.P.~Lilly, J.L.~Reno, P.J.~Burke, and J.P.~Eisenstein, Appl.\ Phys.\ Lett. {\bf 81}, 1627 (2002).

\bibitem{Barnes} W.L.~Barnes, A.~Dereux, and T.W.~Ebbesen, Nature (London) {\bf 424}, 824 (2003).

\bibitem{Shalaev}  {\em Nanophotonics with Surface Plasmons}, edited by V.M.~Shalaev and S.~Kawata (Elsevier, Amsterdam, 2007).

\bibitem{Muravev} V.M.~Muravev, I.V.~Kukushkin, J.~Smet, and K.~von~Klitzing,  Pis'ma Zh.\ Eksp.\ Teor.\ Fiz. {\bf 90}, 216 (2009) [JETP Letters {\bf 90}, 197 (2009)].

\bibitem{Fano} U.~Fano, J.\ Opt.\ Soc.\ Am. {\bf 31}, 213 (1941).

\bibitem{Ritchie} R.H.~Ritchie, Phys.\ Rev. {\bf 106}, 874 (1957).

\bibitem{Stern_surf} E.A.~Stern and R.A.~Ferrell, Phys.\ Rev. {\bf 120}, 130 (1960).

\bibitem{Pitarke} J.M.~Pitarke, V.M.~Silkin, E.V.~Chulkov, and P.M.~Eche\-ni\-que, Rep.\ Prog.\ Phys. {\bf 70}, 1 (2007).

\bibitem{Sondergaard} T.~S{\o}ndergaard, phys. stat. sol. (b) {\bf 244}, 3448 (2007).

\bibitem{Rotenberg} N.~Rotenberg and J.E.~Sipe, Phys.\ Rev.\ B {\bf 83}, 045416 (2011).

\bibitem{Satou} A.~Satou and S.A.~Mikhailov, Phys.\ Rev. B {\bf 75}, 045328 (2007).

\bibitem{Bozhevolnyi} S.I.~Bozhevolnyi and V.~Coello, Phys.\ Rev. B {\bf 58}, 10899 (1998).

\bibitem{Neufeld} J.~Neufeld and R.H.~Ritchie, Phys.\ Rev. {\bf 98}, 1632 (1955).

\bibitem{Echenique} P.M.~Echenique, F.J.~Garc\'ia de Abajo, V.H.~Ponce, and M.E.~Uranga, Nucl.\ Instr.\ and Meth.\ in Phys.\ Res. B {\bf 96}, 583 (1995).

\bibitem{Stern} F.~Stern, Phys.\ Rev.\ Lett. {\bf 18}, 546 (1967).

\bibitem{Horing} N.J.M.~Horing and M.~Yildiz, Phys.\ Lett. A {\bf 44}, 386 (1973).

\bibitem{Chiu} K.W.~Chiu and J.J.~Quinn, Phys.\ Rev. B {\bf 9}, 4724 (1974).

\bibitem{Husimi} K.~Husimi, Prog.\ Theor.\ Phys. {\bf 9}, 381 (1953).

\bibitem{Dmitriev} I.A.~Dmitriev, A.D.~Mirlin, and D.G.~Polyakov, Phys.\ Rev. B {\bf 75}, 245320 (2007).

\bibitem{Zudov} M.A.~Zudov, R.R.~Du, J.A.~Simmons, and J.L.~Reno, Phys.\ Rev. B {\bf 64}, 201311 (2001).

\bibitem{obmen} R.H.~Ritchie and J.C.~Ashley, J.\ Phys.\ Chem.\ Solids {\bf 26}, 1689 (1965).

\bibitem{Ando} T.~Ando, A.B.~Fowler, and F.~Stern, Rev.\ Mod.\ Phys. {\bf 54}, 437 (1982).

\bibitem{Heiblum} M.~Heiblum, E.E.~Mendez, and F.~Stern, Appl.\ Phys.\ Lett. {\bf 44}, 1064 (1984).

\bibitem{Zudov2} M.A.~Zudov, R.R.~Du, L.N.~Pfeiffer and K.W.~West, Phys.\ Rev.\ Lett.  {\bf 90}, 046807 (2003).

\bibitem{Sidorov} Yu.V.~Sidorov, M.V.~Fedoryuk, and M.I.~Shabunin, {\em Lectures on the
Theory of Functions of a Complex Variable} (Nauka, Moscow, 1989) [in Russian].

\bibitem{Chaplik} A.V.~Chaplik, Zh.\ Eksp.\ Teor.\ Fiz. {\bf 62}, 746 (1972) [Sov.\ Phys.\
JETP {\bf 35}, 395 (1972)].

\bibitem{Volkov} V.A.~Volkov and E.E.~Takhtamirov, Zh.\ Eksp.\ Teor.\ Fiz. {\bf 131}, 681 (2007) [JETP {\bf 104}, 602 (2007)].

\bibitem{Shaner} E.A.~Shaner and S.A.~Lyon, Phys.\ Rev. B {\bf 66}, 041402(R) (2002).

\bibitem{Bateman} H.~Bateman, {\em Higher Transcendental Functions}, edited by A.~Erd{\' e}lyi (McGraw-Hill, New York, 1953), Vol. 2.

\bibitem{ChiuLeeQuinn} K.W.~Chiu, T.K.~Lee, and J.J.~Quinn, Surf.\ Sci.\ {\bf 58}, 182 (1976).

\bibitem{Landavshitz} L.D.~Landau and E.M.~Lifshitz, {\em Electrodynamics of Continuous Media} (Pergamon, Oxford, 1984).

\bibitem{Boyd} R.W.~Boyd, {\em Nonlinear Optics} (Academic Press, San Diego, 2003).

\bibitem{Novotny} L.~Novotny, B.~Hecht and D.W.~Pohl, J.\ Appl.\ Phys.\ {\bf 81} 1798 (1997).

\bibitem{Landavshitz2} L.D.~Landau and E.M.~Lifshitz, {\em The Classical Theory of Fields} (Butterworth-Heinemann, Oxford, 1980).

\end{thebibliography}
\end{document}